\documentstyle[preprint,eqsecnum,aps,floats,epsf]{revtex}
\begin{document}

\unitlength=1mm

\def\a{{\alpha}}
\def\b{{\beta}}
\def\e{{\epsilon}}
\def\k{{\kappa}}
\def\l{{\lambda}}
\def\m{{\mu}}
\def\n{{\nu}}

\def\thCM{{\theta_{\text{cm}}}}
\def\o{{\omega}}
\def\th{{\theta}}


\preprint{\vbox{
  \hbox{NT@UW-01-02}
  \hbox{nucl-th/0102040}
}}

\tighten

\title{How a quark-gluon plasma phase modifies the bounds on \\
       extra dimensions from SN1987a}
\author{Daniel Arndt\footnote{arndt@phys.washington.edu}}
\address{Department of Physics, University of Washington,
Seattle, WA 98195-1560. }
\maketitle

\begin{abstract}
The shape of the neutrino pulse from the supernova SN1987a provides
one of the most stringent constraints on the size of large, compact,
``gravity-only'' extra dimensions.
Previously, calculations have been carried out for a newly-born
proto-neutron star with a temperature of about 50~MeV at nuclear matter
density.
It is arguable that, 
due to the extreme conditions
in the interior of the star, matter
might be 
a quark-gluon plasma, where the relevant degrees of freedom
are quarks and gluons rather than nucleons.
We consider an energy-loss scenario 
where seconds after rebounce
the core of the star consists of a hot and dense quark-gluon plasma.
Adopting a simplified model of the plasma
we derive the necessary energy-loss
formulae in the soft-radiation limit.
The emissivity is found to be 
comparable to
the one for nuclear matter
and
bounds on the radius of extra dimensions are similar to 
those found previously from nuclear matter
calculations.
\end{abstract}


\section{Introduction}
Ideas that we live in a world consisting
of $d$ compact extra dimensions in addition
to the usual four infinite dimensions are not new and date back
to the 1930s.
They imply that 
there are Kaluza-Klein~(KK)-modes corresponding to excitations
of ordinary standard model particles in the extra dimensions.
Such modes, however, have not been seen in any collider
experiment so far~\cite{Reviewofparticlephysics}.

Recently, a variation of this concept has been revitalized by 
Arkani-Hamed {\it et al.}~\cite{ADD98,ADD99,AAD98} 
who considered an alternative
picture in which standard model fields are confined to a 
four-dimensional ``brane'' while gravity, 
constituting the dynamics of space-time itself, 
is allowed to propagate into
the whole (4+$d$) dimensional ``bulk''.  
This picture caused some excitement because it provides a natural
solution to the hierarchy problem.  As detailed in~\cite{ADD99},
the extension of the four-dimensional space for gravitons leads to
a ``dilution'' of gravity into the extra dimensions 
and therefore to the smallness of Newton's
gravitational constant $G_N$.  
Matching the Planck mass $M_{\text{Pl}(4)}$ in the four-dimensional
world to the one in the (4+$d$) dimensional world,
$M_{\text{Pl}(4+d)}$, via the relation
\begin{equation}
  M_{\text{Pl}(4)}^2 \sim R^d M_{\text{Pl}(4+d)}^{2+d}
,\end{equation}  
one can bring the Planck mass down from $10^{19}$~TeV
to the weak scale of $~1$~TeV by choosing an appropriate size $R$
of the extra 
dimensions.\footnote{
  A different way of solving the hierarchy problem has been proposed
  by Randall and Sundrum~\cite{RaS99}.  In their model the metric
  is non-factorizable but rather the four-dimensional metric
  is multiplied by an exponential ``warp'' factor.  
  We do not consider this model here.}
For one extra dimension, $R\sim 10^{11}$~m, in this case
gravity---specifically Newton's inverse-square law---would 
be modified on the scale of astronomical distances,
a case clearly ruled out.  
But already for $d=2$, $R\sim 1$~mm,
which is just out of reach to present experimental
measurements of the gravitational force law~\cite{HSH00}.  

Another way of testing these models is to search
for ``beyond the standard model'' physics at the scale
$M_{\text{Pl}(4+d)}$.  Even though one would naively expect a
violation of the standard model at this scale to be of
only gravitational strength, the large number of excited
KK-modes at such high energies compensates for the weak coupling
between these KK-modes and the ordinary matter fields.
Such effects could alter standard model 
predictions due to virtual KK-mode contributions 
or manifest themselves in the form
of massive graviton production~\cite{HLZ99,GRW99,Hew99}. 
It might be possible that such effects can 
be seen in the form of missing energy 
at future colliders operating at the TeV scale
like the CERN LHC~\cite{VaH00,AnB00}.

The most stringent constraints on the size of extra dimensions
come from astrophysical considerations.
Recently it has been pointed out that an overproduction of massive
KK modes in the early universe might result in early matter domination
and therefore a lower age of the universe~\cite{Fai01}.  

Another way to obtain bounds on the size of extra dimensions 
from astrophysics comes from
supernova SN1987a.
Our current theory of supernovae predicts 
the shape and duration of SN1987a's neutrino pulse very well.
Hence, if
some new channel transports too much energy 
from the interior of the supernova then
the current understanding of SN1987a's neutrino signal gets
invalidated.
Raffelt~\cite{Raffeltbook} has quantified the maximum possible 
emissivity for 
a new energy loss process which doesn't conflict with our 
current understanding of the neutrino signal
as $10^{19}$~ergs/g/s.  
Using this criteria there have been several
calculations~\cite{BHK99,CuP99,HPR00} 
to obtain bounds
on the size of the extra dimensions for a newly born proto-neutron star  
assuming that the interior of the star
consists of nuclear matter at a temperature of $\sim 50$~MeV.
A more rigorous approach has been undertaken recently by 
Hanhart {\it et al.}~\cite{HPP01} who performed detailed
simulations of the effect of
exotic radiation on the neutrino signal.

Although the star's inner core might well consist of
nuclear matter, because of the extreme conditions seconds
after core bounce with a density in some regions
possibly up to $\sim 15$ 
times the nuclear matter
density $n_0=0.17~\text{fm}^{-3}$
and a temperature of maybe up to $\sim 100$~MeV, 
matter is near a 
phase transition to a deconfined QCD plasma. 

It is therefore important to investigate how bounds on the size
of extra dimensions change if a scenario is considered 
where, 
in contrast to taking nucleons as the relevant degrees of freedom,
the calculation is carried out
using quarks and gluons 
as degrees of freedom.
However, in the regime we are considering here matter
is not in a state where the density
and temperature are so high that a first order perturbative
calculation would suffice; it is rather in a regime where
deconfinement just sets in and the QCD coupling constant is still 
large ($\a_s\sim 1$).
For this reason we will carry out our calculations for a (unphysical)
regime of very high density (up to $10^7n_0$).  There,
the coupling is weak and perturbative calculations are feasible.
From there we extrapolate down to get the emissivity in 
the physical region.
 
Note that if the temperature falls below a critical temperature $T_c$ of
$\sim 50$~MeV (for a density of a few times nuclear density) 
this very dense plasma most likely undergoes
a phase-transition to a color superconducting phase 
(for a recent review see~\cite{CSCreview}).
Since we do not consider such a scenario here, we assume 
the temperature to be above $T_c$.

The paper is organized as follows:
In Section~\ref{S:KK-loss} 
we calculate the amplitude for the emission of soft gravi- and dilastrahlung
into KK-modes from quark-quark (qq) scattering in a degenerate QCD plasma.
This calculation is fairly general, it applies to soft radiation of 
KK-modes from light relativistic fermions.
Then, in Section~\ref{S:quark-quark}, we consider a simple model of
the plasma consisting of three light quark flavors to estimate
the size of the qq scattering cross sections.
To treat the occurring divergences for colinear
scattering we incorporate
the effects of the surrounding plasma
into the gluon propagator by introducing a cutoff mass.
We do not consider any other many-body effects;
our goal is to obtain an estimate
for the qq scattering cross section,
an exact calculation is presently not feasible.
In Section~\ref{S:emissivity}, we carry out the phase-space
integration to get a formula for numerical calculation
of the emissivity of a gas of weakly interacting quarks.
We also provide an approximation formula for the degenerate limit.
Finally, in Section~\ref{S:results},
we calculate the emissivity and 
the bounds on the size of the $d$ extra dimensions
for the case of $d=2$ and $d=3$ and compare them to the
result from previous nuclear matter calculations.

\section{4D-Graviton, KK-Graviton, and KK-Dilaton Radiation from
         Quark-Quark Scattering}
\label{S:KK-loss}
We assume the $d$ extra dimensions to form a $d$-torus with
radius $R$ so that the KK-graviton and KK-dilation mode expansions
are given by
\begin{equation}
  h^{\mu\nu}(x,y)=\sum_{\vec{j}} h^{\mu\nu,\vec{j}}(x)
                  \exp\left(i\frac{\vec{j}\cdot \vec{y}}{R}\right)
  \quad\text{and}\quad
  \phi_{ab}(x,y)=\sum_{\vec{j}} h_{ab}^{\vec{j}}(x)
                  \exp\left(i\frac{\vec{j}\cdot \vec{y}}{R}\right)
,\end{equation}
respectively.
In addition to these particles there 
emerge massive spin-1 particles from the
KK reduction of the Fierz-Pauli Lagrangian~\cite{HLZ99} 
which decouple from matter and need not be considered 
here.%
\footnote{
Note that while the torodial compactification is conceptually 
simple, it might not be realistic since it doesn't take into account
curvature (``warping'') caused by fields in the bulk and on the brane.
As has been recently investigated by Fox~\cite{Fox00},   
weak warping might increase the emissivity by a factor $\sim 2$
and therefore strengthen the bounds.}

The coupling of the 4D-graviton $g\equiv h$, KK-graviton $h^{\vec j}$, 
and KK-dilaton $\phi^{\vec j}$ to the
energy-momentum tensor $T_{\mu\nu}$ of the matter field
is governed by the Lagrange density
\begin{equation} \label{E:lagrangian}
  {\cal L} 
    = -\frac{\kappa}{2} 
      \left(
      h^{\mu\nu} T_{\mu\nu}
      +\sum_{\vec j}
      \left[ 
        h^{\mu\nu,\vec j} T_{\mu\nu}
        + \sqrt{\frac{2}{3(d+2)}} \phi_{aa}^{\vec j} T^\mu_\mu
      \right]
      \right)
,\end{equation}
where $\kappa$ is defined as $\kappa=\sqrt{16\pi G_N}$ with $G_N$
being Newton's gravitational constant.  
Note that 
in Eq.~(\ref{E:lagrangian}) the massless graviton 
$h_{\mu\nu}$---being 
the zero-mode of the $h^{\mu\nu,\vec j}$ modes---has been 
written separately for clarity.
The energy-momentum tensor $T_{\mu\nu}$ for a free quark
is given by
\begin{equation}
  T_{\mu\nu} 
  = 
  \frac{i}{2}\bar{\psi}
  \left(\overrightarrow{\partial_\mu}-\overleftarrow{\partial_\mu}\right)
  \gamma_\nu\psi
  - g_{\mu\nu}
    \left[\frac{i}{2}\bar{\psi}\gamma^\a
    (\overrightarrow{\partial_\a}-\overleftarrow{\partial_\a})
    \psi-m\bar{\psi}\psi\right]
.\end{equation} 

As we will discuss in the next section, the leading contribution
from gravitational emission for the degenerate quark-gluon plasma
of our model comes from the qq scattering processes
$qq\rightarrow qqX$,
where a low-momentum 4D-graviton ($X=g$), 
KK-graviton ($X=h$), or KK-dilaton ($X=\phi$) is emitted.

\subsection{Amplitude for $qq\rightarrow qqX$ in the Soft Limit}
\begin{figure}[htb]
  \centerline
    {\epsfxsize=4.5in \epsfbox{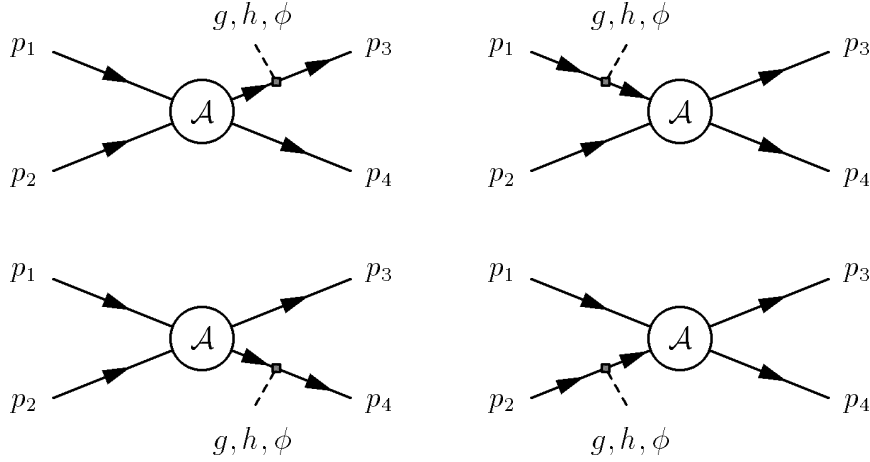}}
    \vskip .2in
    \caption{The leading order diagrams for the process 
          $qq\rightarrow qqX$.
          In the soft limit these
          diagrams are dominant over those where the graviton 
          couples to some internal line.
          ${\cal A}$ is the on-shell 
          amplitude for quark-quark scattering.}
  \label{F:brems-diagrams}
\end{figure}

For a degenerate plasma the participating quarks are near 
the Fermi surface and hence the momentum $\vec{k}$ of the
emitted graviton is small compared to the quark momenta.
In the soft limit the leading diagrams for the process
$qq\rightarrow qqX$
are those where the graviton radiates
off one of the external legs as shown in Fig.~\ref{F:brems-diagrams} 
(bremsstrahlung process).
For small $\vec{k}$ these diagrams
are ${\cal O}(k^{-1})$ while
diagrams where the graviton couples to some internal line
are ${\cal O}(1)$ and therefore suppressed~\cite{Low58,AdD66,HPR00a}.

The amplitude $M_{\mu\nu}$ 
of this process then factorizes into the amplitude ${\cal A}$
(which includes a sum over spins and colors of the incoming
and outgoing quarks) for on-shell
$qq\rightarrow qq$ scattering and the conserved gravitational current 
$T_{\mu\nu}$.
To lowest order in $\vec{k}$, $M_{\mu\nu}$ can be written
as~\cite{HPR00}
\begin{equation} \label{E:m-munu}
  M_{\mu\nu} 
    = \frac{\kappa}{2}
      \sum_{j=1}^4 \eta_j {p_j}_\mu {p_j}_\nu \frac{1}{p_j \cdot k} 
      \,{\cal A}
,\end{equation}
where
\begin{equation}
  \eta_j = 
     \left\{ 
       \begin{array}{cl}
         1  & \text{if } j=1,2 \\
         -1 & \text{if } j=3,4
       \end{array}
     \right.\\
.\end{equation}
Here, the $p_j$ denote the four-momenta of the incoming ($j=1,2$) and
outgoing ($j=3,4$) quarks while $k=(\o,\vec{k})$ is the
graviton's four-momentum.
The result Eq.~(\ref{E:m-munu}) has been calculated by 
Weinberg~\cite{Weinbergbook} using an entirely classical derivation.
This is not unexpected since to lowest order in $\vec{k}$ graphs
where the radiation comes from an internal line are suppressed
and so the radiation from the external legs is the dominant one,
corresponding to classical bremsstrahlung from the gravitational
charges accelerated during the collision.

\subsection{Energy-Loss Formulae}
The energy loss into a KK-mode $\vec{j}$ and into 
the momentum space volume element $d^3k$ due to 4D-graviton, KK-graviton,
and KK-dilation emission is given by
\begin{equation} \label{E:deg}
  d\epsilon_g
  = \omega \frac{d^3 k}{(2\pi)^3 2\omega}
    \sum_\lambda
    \left[M^{\a\beta} \epsilon_{\a\beta}(\vec{k},\lambda)\right]^* 
    M^{\mu\nu} \epsilon_{\mu\nu}(\vec{k},\lambda)
  = \omega \frac{d^3 k}{(2\pi)^3 2\omega}
    \left(
      {M^{\lambda\nu}}^* M_{\lambda\nu}
      - \frac{1}{2}\left|M^\lambda_{\phantom{\lambda}\lambda}\right|^2
    \right)
,\end{equation}
\begin{equation} \label{E:deh}
  d\epsilon^{\vec{j}}_h
  = \omega \frac{d^3 k}{(2\pi)^3 2\omega}
    \sum_{\lambda=1}^5
    \left[M^{\a\beta} 
           \epsilon^{\vec{j}}_{\a\beta}(\vec{k},\lambda)\right]^* 
    {M^{\mu\nu} \epsilon^{\vec{j}}_{\mu\nu}(\vec{k},\lambda)}
,\end{equation}
and
\begin{equation} \label{E:dephi}
  d\epsilon^{\vec{j}}_\phi
  = \frac{2}{3(d+2)}
    \omega \frac{d^3 k}{(2\pi)^3 2\omega}
    M_\a^{\phantom{\a}\a} M_\beta^{\phantom{\beta}\beta}
    \sum_{\lambda=1}^{n(n-1)/2} 
    \left[e^{\vec{j}}_{ii}(\vec{k},\lambda)\right]^*
    e^{\vec{j}}_{jj}(\vec{k},\lambda)  
  = \frac{d^3 k}{2(2\pi)^3}
    M_\a^{\phantom{\a}\a} M_\beta^{\phantom{\beta}\beta}
    \frac{2(d-1)}{3(d+2)}
,\end{equation}
respectively.
Here, $\epsilon_{\mu\nu}$, $\epsilon^{\vec{j}}_{\mu\nu}$, 
and $e^{\vec{j}}_{\mu\nu}$
are the polarization tensors
of the 4D-graviton, KK-graviton, and KK-dilaton, respectively
\cite{HLZ99}.

Utilizing the conservation of the energy-momentum tensor,
$k^\mu M_{\mu\nu}=0$, $M_{\mu\nu}$ can be expressed in terms of
the purely spacelike components as
\begin{equation}
  M_{0i} = -\frac{k^j}{\omega}M_{ji},\quad
  M_{00} = \frac{k^i k^j}{\omega^2}M_{ji}
\end{equation}
or, upon defining
\begin{equation}
  W_{\mu\nu ij}
  = \eta_{i\mu}\eta_{j\nu}
    - \eta_{i \mu}\eta_{0 \nu} \frac{k_j}{\omega} 
    - \eta_{0 \mu}\eta_{j \nu} \frac{k_i}{\omega}
    + \eta_{0 \mu}\eta_{0 \nu} \frac{k_i k_j}{\omega^2}
,\end{equation}
be written as
\begin{equation}
  M_{\mu\nu} = W_{\mu\nu ij}M^{ij}
.\end{equation}
Using this notation 
the energy loss formulae in Eqs.~(\ref{E:deg})--(\ref{E:dephi}) 
can be written solely in
terms of the space-space components of the tensor $M_{\mu\nu}$ as
\begin{equation} \label{E:de_x}
  d\epsilon^{\vec{j}}_X
  = \frac{dk k^2}{4\pi^2} \frac{d\hat{k}}{4\pi} 
    M_{ij} \Lambda_X^{ijkl} M_{kl}
,\end{equation}
where the $\Lambda_X^{ijkl}$ are given by
\begin{equation}
  \Lambda_g^{ijkl}
   = \delta^{ik}\delta^{jl}
     - \frac{1}{2}\delta^{ij}\delta^{kl}
     - \frac{1}{2k^2}
       \left(4k^jk^l\delta^{ik}-k^kk^l\delta^{ij}-k^ik^j\delta^{kl}\right)
     + \frac{1}{2k^4}k^ik^jk^kk^l 
,\end{equation}
\begin{eqnarray}
  \Lambda_h^{ijkl}
   &=& W^{\mu\nu ij}B_{\mu\nu\a\beta}W^{\a\beta kl} \nonumber \\
   &=& \left(
         \delta^{ik}\delta^{jl} + \delta^{il}\delta^{jk}
         -\frac{2}{3}\delta^{ij}\delta^{kl}
       \right)
       + \frac{4}{3\omega^4}k^ik^jk^kk^l \nonumber \\
    && - \frac{1}{\omega^2}\left(
                             \delta^{ik}k^jk^l + \delta^{jk}k^ik^l
                             + \delta^{il}k^jk^k + \delta^{jl}k^ik^k
                             -\frac{2}{3}(k^ik^j\delta^{kl}+k^kk^l\delta^{ij})
                           \right)
,\end{eqnarray}
and
\begin{equation}
  \Lambda_\phi^{ijkl}
   = W_{\a\phantom{\a}}^{\phantom{\a}\a ij} 
     W_{\beta\phantom{\beta}}^{\phantom{\beta}\beta kl}
     \frac{2(d-1)}{3(d+2)}
   = \left(\delta^{ij}\delta^{kl}
           +\frac{1}{\omega^2}\left[\delta^{ij}k^kk^l+\delta^{kl}k^ik^j\right]
           +\frac{1}{\omega^4}k^ik^jk^kk^l
     \right)
     \frac{2(d-1)}{3(d+2)}
.\end{equation}
The tensor $B_{\mu\nu\a\beta}$, defined by the spin sum of
the KK-graviton polarization tensors 
\begin{equation}
  \sum_{\l=1}^5 \epsilon_{\mu\nu}^{\vec{j}}(\vec{k},\l)
  \left[\epsilon_{\rho\sigma}^{\vec{j}}(\vec{k},\l)\right]^*
  =
  B_{\mu\nu\rho\sigma}^{\vec{j}}(\vec{k})
,\end{equation}
is given in~\cite{HLZ99,GRW99}. 

We will now choose the center-of-momentum (COM)
frame as our reference frame 
where the 4-momenta of the two incoming
quarks are given by $(p^0,\vec{p})$ and $(p^0,-\vec{p})$ and  
those of the outgoing quarks by 
$({p'}^0,\vec{p'})$ and $({p'}^0,-\vec{p'})$.
The COM scattering angle $\thCM$ is defined by
$\vec{p}\cdot\vec{p'}=p p' \cos{\thCM}$.
We also assume the quarks to have equal mass.

If the momenta of the scattering particles 
are much smaller than their 
rest mass $M$%
---%
as is the case in soft radiation from nucleon-nucleon scattering%
---%
it can be neglected in the denominator in Eq.~(\ref{E:m-munu})
and $M_{\mu\nu}$ is
simply given by~\cite{HPR00}
\begin{equation}
  M_{ij}=\frac{\k}{M\o}\left(p_ip_j-p'_ip'_j\right)\,{\cal A}
.\end{equation}
In our case, however, this approximation is inappropriate 
since the
quark mass is very small compared to the quark Fermi momenta,
we therefore write Eq.~(\ref{E:m-munu}) as
\begin{equation} \label{E:M-ij}
  M_{ij}
  = \frac{\kappa}{2}
    \left(
      \frac{p_i p_j}{p^0\omega-\vec{p}\cdot\vec{k}}
      +\frac{p_i p_j}{p^0\omega+\vec{p}\cdot\vec{k}}
      -\frac{p'_i p'_j}{p'^0\omega-\vec{p'}\cdot\vec{k}}
      -\frac{p'_i p'_j}{p'^0\omega+\vec{p'}\cdot\vec{k}}
    \right){\cal A}
.\end{equation}
The $\int{d\hat{k}}/{4\pi}$ integration in Eq.~(\ref{E:de_x}) 
can then 
be carried out analytically for the three cases.
For soft graviton radiation 
the quark momenta are assumed to lie on the Fermi surface and 
it is justified to assume
$|\vec{p}|\approx|\vec{p'}|$.  We can then
approximate the momenta $p$ and $p'$ by
\begin{equation}
  \bar{p}^2=\frac{p^2+{p'}^2}{2}
.\end{equation}
Replacing $p$ and $p'$ with $\bar{p}$ the formulae simplify 
considerably and the 
$\int{d\hat{k}}/{4\pi}$ integration
can be written as
\begin{equation} \label{E:g-define}
  \int \frac{d\hat{k}}{4\pi}  M_{ij} \Lambda_X^{ijkl} M_{kl}
  =
  \frac{\k^2\bar{p}^2}{\o k} g_X(x,y,\thCM)
  \left|{\cal A}\right|^2
\end{equation}
in terms of ${\cal A}$ and the dimensionless functions $g_X$
which are given in the Appendix.
Here we have also defined $x=m_{\vec{j}}/\o$ and 
$y={\bar{p}}^0/\bar{p}={\sqrt{{\bar{p}}^2+M^2}}/\bar{p}$.

Note that for massless quarks one of the denominators in 
Eq.~(\ref{E:M-ij})
becomes zero if the momentum of that quark is parallel to $\vec{k}$
and one might think that this term becomes infinite.
However, as already noted in Ref.~\cite{Weinbergbook} for the 
case of the massless 4D-graviton, 
this singularity is only spurious and gets canceled
by $\Lambda_X^{ijkl}$ in the numerator so that $g_X$ is 
finite not only for $X=g$, but also for $X=h$ and $X=\phi$.

We can then employ the energy-momentum
relation for the KK-mode $\vec{j}$ and
write the formula for the energy loss per unit frequency
interval into the given KK-mode of mass 
$m_{\vec{j}}^2=\o^2-k^2$ as
\begin{equation} \label{E:deX-domega}
  \frac{d\epsilon_X^{\vec{j}}}{d\o}
  =
  \frac{\o k}{4\pi^2}
  \int \frac{d\hat{k}}{4\pi} M_{ij} \Lambda_X^{ijkl} M_{kl}
  =
  \frac{\k^2\bar{p}^2}{4\pi^2} g_X(x,y,\thCM)\left|{\cal A}\right|^2
.\end{equation}
This result is general as it applies to soft radiation from
any light relativistic
fermions with equal mass, 
not just to the quarks we are about to consider.

\section{A Simple Plasma Model}
\label{S:quark-quark}
After bounce the homologous core of the star undergoing a 
supernova explosion gets compressed to a very high density.
Previous calculations \cite{BHK99,CuP99,HPR00}
assumed a density of $1-3$ times nuclear matter density 
$n_0=0.17$~fm
and a temperature of $T\approx 30-80$~MeV corresponding 
to a (non-relativistic)
nucleon chemical potential of 
$\mu_N\approx 50-100$~MeV. 

Under such extreme conditions the star's core not only consists
of free neutrons and protons but these degrees of freedom are
highly excited and near a phase transition to a deconfined
quark-gluon plasma.
It is therefore worthwhile to investigate
how the graviton emissivity and hence the bounds on extra
dimensions change if one carries out the calculations in this
regime.
Our intention here is to investigate graviton radiation into 
extra dimension from such a QCD plasma.

The physics of a QCD plasma under such violent conditions is particularly
rich.  In addition to quarks of different flavors, a variety of
other particles are present and being produced constantly, such as
electrons, positrons, neutrinos, and collective excitations like 
plasmons.  
The properties of the quark-gluon plasma are far from being 
completely understood so that we do not even attempt to take the
whole richness of matter under such extreme conditions into
account here.
Instead, we assume a rather simply model for the plasma (not unlike
the one used in Ref.~\cite{Iwa82}) where, 
in addition to electrons and neutrinos, 
only $u$, $d$, and $s$
quarks are assumed to be present.
The possible weak reactions in the plasma are then
\begin{equation}
  d\leftrightarrow u+e^-+\n_e^- \quad\text{and}\quad
  s\leftrightarrow u+e^-+\n_e^-
.\end{equation}
At thermal equilibrium the chemical potentials
have to add up to zero.  For a low temperature $T$
this requirement reads
\begin{equation}
  \m_d=\m_u+\m_e+\m_\n \quad\text{and}\quad
  \m_s=\m_u+\m_e+\m_\n
.\end{equation}
In addition, we want to satisfy charge neutrality, thereby imposing
the constraint
\begin{equation}
  0=Q=\frac{2}{3}n_u-\frac{1}{3}n_d-\frac{1}{3}n_s-n_e
.\end{equation}
These three conditions can be simultaneously satisfied by choosing
\begin{equation}
  n_u=n_d=n_s=\frac{n_q}{3}=n_B \quad\text{and}\quad
  n_e=n_\n=0
,\end{equation}
where $n_q$ is the quark number density and
$n_B$ denotes the baryon number density.
We do not consider dynamical production
of these or any heavier quark flavors.   
As an additional simplification
we take the mass of the $u$, $d$, and $s$ quarks to be zero.

The main contribution to graviton radiation from this plasma
comes from the bremsstrahlung reaction $qq\rightarrow qqX$
($X=g,h,\phi$) 
where a graviton is radiated from an external quark leg.  
This reaction is of order
${\cal O}(\k\a_s)$ where
$\a_s$ is the QCD coupling constant.  As we will see below,
$\a_s$ is rather large and ${\cal O}(1)$.
Moreover, we get contributions from processes involving 
dynamical produced electrons ($e$) and positrons ($p$), like
$ee\rightarrow eeX$ or $ep\rightarrow epX$.  These reactions
are suppressed compared to
$qq\rightarrow qqX$ by a factor of $\a_{\text{QED}}/\a_s$, where
$\a_{\text{QED}}=1/137$ is the fine-structure constant.
More contributions come from graviton-production reactions
like $q\rightarrow qXX$.  These, however,
are ${\cal O}(\k^2)$ and therefore highly suppressed.
We also assume that because of $T\ll \mu_q$ the number density of
gluons is much smaller than that of quarks and
therefore we do not consider the reaction $qg\rightarrow qX$.
Note that the inclusion of any KK-modes generated by channels
we neglect
will only increase the emissivity and strengthen the bounds
obtained below.
  
The baryon number density $n_B$ 
in the central core region of a stable neutron star
is $\sim 1-15$ times the nuclear matter density
$n_0=0.17~\text{fm}^{-3}$,
depending on the nuclear matter equation of state 
used (compare Ref.~\cite{Iwa82} and values cited therein).
Assuming quark matter, the (relativistic) 
quark chemical potential $\mu_q$
can be related to the quark density $n_q$ using the formula
\begin{equation}
  n_q=6\frac{1}{2\pi^2}
      \int_0^{\infty}d\e\,\e^2 N_q(\e)
\end{equation}
where $N_q(\e)=(\exp[(\e-\mu_q)/T]+1)^{-1}$ is the 
mean occupation number and $\e$ is the (relativistic) quark energy.
Choosing $n_q$ and $T$ we can then determine
$\mu_q$; 
for $T=50$~MeV and $n_q=10-20 n_0$ we find
$\mu_q\approx 500-600$~MeV, which is in good agreement with
other calculations~\cite{Iwa82}.
Therefore quark matter has $\mu_q/T\gg 1$ which confirms our
assumption that it is degenerate.

Furthermore, using $\mu_q$ and $T$ we can calculate
the average momentum square $\langle p^2\rangle$ carried by the quarks using
\begin{equation}
  \langle p^2\rangle
  =\frac{\int d^3p\,p^2N_q(\e)}{\int d^3p\,N_q(\e)}
.\end{equation}
For typical parameters for the plasma 
($n_q\approx 10-20 n_0$, $T\approx 50$~MeV) we calculate
the average momentum carried by the quarks to be 
$p\approx 420-520$~MeV which is just  about the order of energy
at which the strong interaction becomes strong and the 
QCD coupling constant $\a_s$ is expected to be rather large.
Of course, this is eventually a sign
that matter under these conditions is at
the phase transition to the QCD phase.
Therefore this regime is particularly difficult to treat because on the one
hand
nucleonic degrees of freedom are no longer the 
proper degrees of freedom while on the other hand
perturbative QCD calculations are 
notoriously difficult.

For this reason we will evaluate the emissivity for
unphysically high
densities of  $n_q\sim100-10^7 n_0$ and extrapolate from this region
down to the 
physical regime.  At such high densities 
the momentum carried by the quarks 
is $p\sim 1-40$~GeV.
This being in the perturbative regime, we can     
estimate the strength
of the QCD coupling constant $\a_s$ employing the leading-log 
formula for the
running coupling [derived from the $\b$-function to 
${\cal O}(\a_s^2)$]
\begin{equation} \label{E:as}
  \a_s=\frac{g^2}{4\pi}
      =\frac{12\pi}{(33-2N_F)\ln\frac{\langle p^2\rangle}{\Lambda^2}}
\end{equation}
which gives $\a_s \approx0.48-0.13$---small enough for perturbative
calculations.
To check the effects of higher order correction we also use the
formula 
\begin{equation} \label{E:ashigherorder}
  \a_s=\frac{g^2}{4\pi}
      =\frac{4\pi}{\b_0 u}
       \left(1-\frac{2\b_1}{\b_0^2}\frac{\ln u}{u}
             +\frac{4\b_1^2}{\b_0^4 \ln^2u}
               \left[\left(\ln u-\frac{1}{2}\right)^2
                       +\frac{\b_2\b_0}{8\b_1^2}-\frac{5}{4}
                \right]
       \right)
\end{equation}
derived from the $\b$-function to 4th order in $\a_s$.
Here
\begin{equation}
  \b_0=11-\frac{2}{3}N_F,\quad
  \b_1=51-\frac{19}{3}N_F,\quad
  \b_2=2857-\frac{5033}{9}N_F+\frac{325}{27}N_F^2,\text{ and }
  u=\ln\frac{\langle p^2\rangle}{\Lambda^2}
,\end{equation}
where $\Lambda\approx 200$~MeV is taken as the QCD breakdown scale and
$N_F=3$ is the number of flavors.
The formula in Eq.~(\ref{E:ashigherorder})
changes $\a_s$ by about $18-37\%$ compared to Eq.~(\ref{E:as}).

It is instructive to compare the number of nucleons $N_N$ and
quarks $N_q$ 
participating in scattering for the two cases of
nucleon-nucleon scattering 
(for simplicity we assume here neutron matter)
and qq scattering in the degenerate limit.
For low temperature the baryon number density $n_B$ 
is approximately
proportional to $p_F^3$.
For nucleons 
$n_B\approx p_F^3/(3\pi^2)$,
while for quarks  
$n_B\approx p_F^3/\pi^2$.  
The number of particles $N_N$ [$N_q$] 
participating in scattering is roughly 
proportional to the momentum-space volume of a shell of the 
Fermi sphere with radius $p_F$ and
thickness $dE\approx T$
leading to
$N_N\approx 8\pi p_F^2\,dp$ [$N_q\approx 72\pi p_F^2\,dp$] 
in the nucleon [quark] case.  
Making use of the energy-momentum relations
$p\,dp=M\,dE$ [$p\,dp=p\,dE$] this becomes
$N_N\approx \pi(3\pi^5)^{1/3}n_B^{1/3}MT$
[$N_q=72\pi^{7/3}n_B^{2/3}T$].
Using $M\approx 940$~MeV the ratio of $N_N$ and $N_q$ for a
density of $n_B=10n_0$ 
is
\begin{equation}
  \frac{N_N}{N_q} \approx 0.3
\end{equation}
and hence, despite the fact that the quarks have a higher number
of degrees of freedom, the number of particles participating 
in scattering is not much different in the two cases.

\subsection{Quark-Quark Scattering Cross Sections}
\begin{figure}[htb]
  \centerline
    {\epsfxsize=5.5in \epsfbox{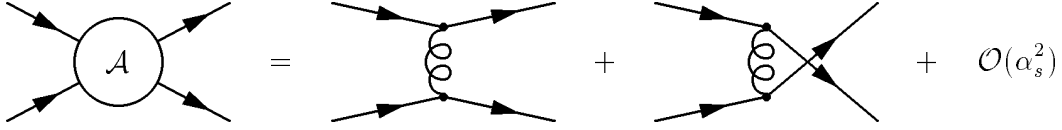}}
  \vskip .2in
  \caption{The lowest order diagrams for scattering of 
           quarks of the same flavor.}
  \label{F:one-gluon}
\end{figure}

The matrix elements
for the scattering of quarks
to first order in $\a_s$
can be calculated from the
one-gluon exchange diagrams
and are given in~\cite{CKR77}%
\footnote{
  These matrix elements, however, include an averaging [summing]
  over initial [final] spin and color states.
  Since for the calculation of the emissivity 
  we need to sum over spin and color of both the initial and final states, 
  we have to include an additional factor of 36 in
  the squared matrix elements.
}.
For quarks of different flavor the squared matrix element
can be written as
\begin{equation} \label{E:A-qq}
  \left|{\cal A}_{qq'}\right|^2 
  = 256 \pi^2 \a_s^2 \frac{s^2+u^2}{t^2}
\end{equation}
and for quarks of the same flavor as
\begin{equation} \label{E:A-qqd}
  \left|{\cal A}_{q q}\right|^2 
  = 256 \pi^2 \a_s^2 
    \left(
      \frac{s^2+u^2}{t^2}+\frac{s^2+t^2}{u^2}
      -\frac{2}{3}\frac{s^2}{u t}
    \right)
\end{equation}
(see Fig.~\ref{F:one-gluon}).
Here $s$, $t$, and $u$ are the usual Mandelstam variables
given in the COM frame for zero quark mass
by $s=4p^2=4{p'}^2$, $t=-2p^2(1-\cos\thCM)$,
and $u=-2p^2(1+\cos\thCM)$,
where $\thCM$ is the COM scattering angle.

Note that similar diagrams exchanging a photon instead of a gluon 
are suppressed
by a factor of $\a_{\text{QED}}/\a_s$ and need not
be considered.

The matrix elements in Eqs.~(\ref{E:A-qq}) and (\ref{E:A-qqd})
can become highly singular because of the
$t^2$ and $u^2$ in the denominators.  The momentum $q$ transfered by
the exchanged gluon vanishes as $\thCM\rightarrow 0$ or
$\thCM\rightarrow \pi$, which corresponds to 
colinear
scattering.  
On the other hand, the energy-momentum tensor $M_{\mu\nu}$ also 
vanishes for colinear scattering
and one might think that this would render the singularity
spurious.
However, this is not the case. An analysis of the 
functions $g_X$ shows that  
for $\thCM\approx0$ [$\thCM\approx\pi$] $g_X$ behaves
like $\sim\thCM^2$ [$\sim(\pi-\thCM)^2$] 
while $\left|{\cal A}_{qq}\right|^2$ and 
$\left|{\cal A}_{qq'}\right|^2$ behave like 
$\sim 1/\thCM^4$ [$\sim 1/(\pi-\thCM)^4$].

\subsection{Curing the Divergences}
In the case of electron-electron scattering in an electron gas
the divergences that one encounters
are of a similar origin and are 
usually taken care of by including 
effects which screen the bare particles arising from
the interaction of the electrons and photons with 
the surrounding matter. 
We will employ, without justification,  
a similar procedure in the present case of a
QCD plasma.

The bare gluon propagator $D_{\mu\nu}$ is related to the
gluon self-energy $\Pi_{\mu\nu}$ by Dyson's equation
\begin{equation}
  D_{\mu\nu}^{-1}=g_{\mu\nu}(\o^2-q^2)+\Pi_{\mu\nu}
.\end{equation} 
Following Ref.~\cite{HP93}, the matrix element for scattering of like quarks
for small momentum transfer is equal to the one for
unlike quarks and can be written as
\begin{equation} \label{E:screening}
  \left|{\cal A}_{qq'}\right|^2
  =128\pi^2\a_s^2
   \left|
     \frac{1}{q^2+\Pi_l^2}-\frac{(1-x^2)\cos{\phi}}{q^2-\o^2+\Pi_t}
   \right|^2
,\end{equation}
where $\Pi_l$ and $\Pi_t$ are the longitudinal and transverse
gluon polarization functions given for $q\ll\mu_q$ by
\begin{equation}\label{E:pi-t}
  \Pi_l=q_D^2\left(1-\frac{x}{2}\ln{\frac{x+1}{x-1}}\right)
  \quad\text{and}\quad
  \Pi_t=q_D^2\left(\frac{x^2}{2}+\frac{x(1-x^2)}{4}\ln{\frac{x+1}{x-1}}\right)
.\end{equation}
Here $x=\o/q$ and 
$q_D$ is the Debye wave number for cold quark matter of
three flavors given by
$q_D^2=6\a_s\mu_q^2/\pi^2$.
Note that for quarks of equal mass in the COM frame $\o=0$ and
therefore $\Pi_t=0$; the transverse part of the 
propagator is still unscreened, at least for $\Pi_t$ calculated to
zeroth
order in $x$.  
Since $T\ll\mu$ the participating
quark's
momenta are near the Fermi surface and hence $\o\simeq 0$, independent
of the chosen reference frame.
The problem of an unscreened transverse gluon propagator can be solved
in some cases by retaining ${\cal O}(x)$ terms in 
Eq.~(\ref{E:pi-t}) and
is related to frequency-dependent screening~\cite{LeBellacbook}.

\begin{figure}[htb]
  \centerline
    {\epsfxsize=4.0in \epsfbox{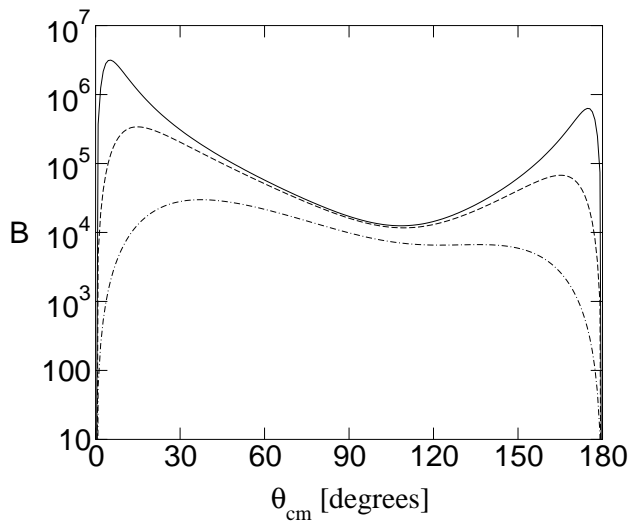}}
  \caption{The function $B(\thCM)$ for $n_q=10^5n_0$
           and for different 
           cutoff masses 
           $m_D/3$~MeV (solid line),
           $m_D$~MeV (dashed line), and
           $3m_D$~MeV (dot-dashed line) with $m_D=2250$~MeV.
           The typical COM momentum is 8400~MeV and $\a_s$
           is set to 1.}
\label{F:qq-xsec}
\end{figure}

Because of these complications 
we adopt here a much simpler method
to incorporate the complicated screening effects.  
We will render the amplitudes in Eqs.~(\ref{E:A-qq}) and (\ref{E:A-qqd})
finite by including a Debye screening mass $m_D$, 
by substituting
\begin{equation}
  t\rightarrow t-m_D^2 \quad\text{and}\quad 
  u\rightarrow u-m_D^2
\end{equation}
for $u$ and $t$ in the denominator.
For a small coupling constant the Debye mass to lowest order is
given by
\begin{equation} \label{E:m_D}
  m_D^2=g^2\left(\frac{3}{2}T^2
                             +\frac{1}{2\pi^2}\sum_f{\mu_f^2}\right)
\end{equation}
which makes it about $60-20\%$ of a typical momentum transfer
for densities $n_q=100-10^7n_0$.
To consider higher order corrections to $m_D$ and
to account for other many-body effects in the QCD plasma at
these high densities we carry out the calculation for the
emissivity not only using the value $m_D$ from Eq.~(\ref{E:m_D}) but
also the values $m_D/3$ and $3m_D$, thereby spanning about an order 
of magnitude in the cutoff mass.
As an example we show in
Fig.~\ref{F:qq-xsec}
the dimensionless function $B(\thCM)$ defined by
\begin{equation}
  B(\thCM)
  =
  \left(\left|{\cal A}_{qq'}\right|^2
  +\frac{\left|{\cal A}_{qq}\right|^2}{4}\right)\sin^2{\thCM}
,\end{equation}
which enters the formula for the emissivity
for a varying cutoff mass $m_D$.  
The factor $\sin^2{\thCM}$ in $B(\thCM)$
is an approximation for 
$\int_0^1 dx\,x^{d-1} g_X(x,1,\thCM)$ as explained in the 
next section.
$B(\thCM)$ has been calculated for a plasma at the density
$n_q=10^5n_0$ where a typical COM momentum is about 8400~MeV and
$m_D\approx 2250~MeV$.
The variation of $m_D$ of about one order of magnitude
causes a variation in $B(\thCM)$ of
about 1-2 orders of magnitude.

Other possible ways to regulate these divergences could be to introduce
a cutoff of order $m_D$ for the gluon-momentum integration
or to cut off the integration over $\thCM$ at some small angle $1/\Lambda$
when calculating observables.  
The latter technique was used by Weinberg~\cite{Weinbergbook} 
to calculate the power produced in gravitational radiation from
the hydrogen plasma in the sun.

\section{Gravitational Emissivity of a Gas of Relativistic
         Fermions}
\label{S:emissivity}
In order to calculate 
the emissivity due to graviton radiation from the two-body 
scattering reaction $qq\rightarrow qqX$ 
in a gas of relativistic quarks we have to integrate over the
phase space of incoming ($\vec{p_1}$, $\vec{p_2}$) 
and outgoing ($\vec{p'_1}$, $\vec{p'_2}$) quarks 
as well as that of the emitted gravitational particle ($\vec{k}$).  
The formula for the emissivity is then given by
\begin{eqnarray} \label{E:emissivity}
  \frac{d{\cal E}_X}{dt} 
    &=&
   \sum_{\vec j} \int d\o
   \int\left[
         \prod_{i=1..2}
         \frac{d^3p_i}{(2\pi)^32E_i}\frac{d^3p_i'}{(2\pi)^32E'_i}
       \right]
   f_1 f_2 (1-f_1')(1-f_2')  \nonumber \\
   &&\times (2\pi)^4\delta^4(p_1+p_2-p_1'-p_2'-k) 
          \frac{d\e_X^{\vec{j}}}{d\o}
.\end{eqnarray}
Here, again, the $X$ stands for 
the type of gravitational radiation, massless 4D-gravitons 
($X=g$),
massive KK-gravitons($X=h$), or massive KK-dilatons ($X=\phi$).  
The functions $f$ are Pauli blocking factors given by
$f_i=(\exp[(E_i-\m)/T]+1)^{-1}$,
furthermore $E_i=\left|\vec{p_i}\right|$ for massless quarks.
The summation over the KK-modes $\vec{j}$ doesn't apply
to the massless 4D-graviton, obviously.

This formula applies for scattering of differently flavored quarks.
A symmetry factor of $1/4$ would have to be included 
for scattering of quarks having the same flavor
to account for identical
particles in the initial and final states.  
We will consider this symmetry factor at the beginning 
of the next section.

In the soft-radiation limit we can neglect the graviton momentum
$\vec{k}$ in the momentum-conserving delta function.  
Furthermore, by introducing the total momentum
$\vec{P}=\vec{p_1}+\vec{p_2}=\vec{p'_1}+\vec{p'_2}$
and relative initial and final momenta
$\vec{p}=(\vec{p_1}-\vec{p_2})/{2}$ and
$\vec{p'}=(\vec{p'_1}-\vec{p'_2})/{2}$
we can reduce the number of integrations by exploiting spherical
symmetry and momentum conservation to get,
after inserting Eq.~(\ref{E:deX-domega}) for the massless quark case
($y=1$),
\begin{eqnarray}
  \frac{d{\cal E}_X}{dt} 
   &=&
   \frac{\k^2}{2^{11}\pi^8}
   \sum_{\vec j} \int d\o
   \int_{-1}^1d\cos\th \int_{-1}^1d\cos\th'
   \int_0^\infty dP\, P^2 \int_0^\infty dp\, p^2 \int_0^\infty dp'\, {p'}^2
   \int_0^{2\pi} d\phi \nonumber \\
   &&\times
   f_1 f_2 (1-f_1')(1-f_2')   \nonumber \\
   &&\times 
   \frac{1}{E_1E_2E'_1E'_2}
   \delta(E_1+E_2-E_1'-E_2')
   \bar{p}^2 g_X(x,1,\thCM)
   \left|{\cal A}(\bar{p},\thCM)\right|^2
.\end{eqnarray}
Here we have also used the COM frame for the amplitude ${\cal A}$.
The angles $\th$ and $\th'$ are defined by 
$\vec{p}\cdot\vec{P}=pP\cos\th$ and
$\vec{p'}\cdot\vec{P}=p'P\cos\th'$, while $\phi$ is the angle 
between the projections of $\vec{p}$ and $\vec{p'}$ on the
$\hat{P}$-plane, that is
\begin{equation}
  \cos\thCM=\cos\phi\sin\th\sin\th'+\cos\th\cos\th'
.\end{equation} 

If the mass splitting of the KK-modes becomes comparable to
the experimental energy resolution---which is true 
for about $d\leq 6$---the summation over the modes $\vec{j}$ can be 
approximated by an integration:
\begin{equation} \label{E:KK-summe}
  \sum_{\vec{j}}
  \longrightarrow
  R^d\Omega_d\int_0^\o dm\,m^{d-1}=R^d\Omega_d\o^d\int_0^1 dx\,x^{d-1}
,\end{equation}
where $d$ is the number of extra dimensions and
$\Omega_d$ is the surface area of the $d$-dimensional
sphere given by
$\Omega_d={2\pi^{d/2}}/{\Gamma(d/2)}$.

To further simplify the integration
we make another change of variables to the initial and
final COM total energies $E_T$ and $E'_T$ given by
\begin{equation}
  E_T=E_1+E_2
     =\sqrt{M^2+\frac{P^2}{4}+p^2+Pp\cos\th}
      +\sqrt{M^2+\frac{P^2}{4}+p^2-Pp\cos\th}
\end{equation}
and a similar expression for $E'_T$.  
The Jacobian ${\cal J}$ of the variable transformation 
$(p,p')\rightarrow(E_T,E'_T)$ is given by
\begin{equation}
  {\cal J}
  =\frac{\partial p }{\partial E_T }\frac{\partial p'}{\partial E'_T}
  =\left\{
   \frac{{E_T}^4-P^2(2{E_T}^2-4M^2-P^2)\cos^2{\th}}
        {2\sqrt{{E_T}^2-4M^2-P^2}(E_T^2-P^2\cos^2{\th})^{3/2}}
   \right\}
   \times \Biggl\{(E_T\rightarrow E'_T),(\th\rightarrow\th')\Biggr\}
.\end{equation}
After interchanging the $E_T$ and
$E'_T$ integrations and shifting the integration variables
the final result for KK-gravitons and KK-dilatons becomes
\begin{eqnarray} \label{E:emissivity-final}
  \frac{d{\cal E}_X}{dt} 
   &=&
   \frac{G_N}{64\pi^7}
   R^d\Omega_d
   \int_0^{\pi}d\th \int_0^{\pi}d\th'
   \int_0^\infty dP \int_0^\infty d\a \int_0^\infty d\a'\,
   \sin\th\sin\th' P^2 \a^d {\cal J} p^2{p'}^2 \nonumber \\
   &&\times
   f_1 f_2 (1-f_1')(1-f_2')  
   \frac{{\bar{p}}^2}{E_1E_2E'_1E'_2}
   \int_0^{2\pi} d\phi
   \left|{\cal A}(\bar{p},\thCM)\right|^2
   \int_0^1dx\,x^{d-1}g_X(x,1,\thCM)
,\end{eqnarray}
where we have defined
\begin{equation}
  \a'=E_T'-2\sqrt{M^2+\frac{P^2}{4}} \quad\text{and} \quad
  \a=\o=E_T-E_T'
.\end{equation}
Note that only the function $g_X$ depends on 
$x=m_{\vec{j}}/\o$.
Also, the integration over $\phi$ is hidden via $\thCM$ in the function 
$g_X$ and therefore nontrivial.

\subsection{Approximation Formula for Degenerate Quark Matter}
Quark matter with a baryon number density
$n_B\sim 10n_0$ has at a temperature $T\sim 50$~MeV
a quark chemical potential  
$\mu_q\sim 500$~MeV and is highly
degenerate, $\mu_q/T\gg 1$.  At higher densities matter is 
even more degenerate.
Therefore one can assume that radiation only arises
from scattering involving quarks near the Fermi surface in the
initial and final states.  
Assuming soft radiation the energy of the 
quarks in Eq.~(\ref{E:emissivity}) can be set to
$p_F$ causing a decoupling of the energy and angular integrations:
\begin{eqnarray}
  \frac{d{\cal E}_X}{dt} 
   &=&
   \frac{p_F^4}{2^{12}\pi^8}
   \int d\o  
     \int dE_1\,dE_2\,dE_1'\,dE_2'
     f_1 f_2 (1-f_1')(1-f_2')
     \delta(E_1+E_2-E_1'-E_2'-\o)  \nonumber \\
   &&\times 
   \sum_{\vec j}
   \int d\Omega_1\,d\Omega_2\,d\Omega_1'\,d\Omega_2'
   \delta(\vec{p_1}+\vec{p_2}-\vec{p_1'}-\vec{p_2'})
   \frac{d\e_X^{\vec{j}}}{d\o}
.\end{eqnarray}
The energy integral is well known~\cite{MoP62,BaP78} and given by
\begin{eqnarray}
  \int dE_1\,dE_2\,dE_1'\,dE_2'
  f_1 f_2 (1-f_1')(1-f_2')
  \delta(E_1+E_2-E_1'-E_2'-\o)
  =
  \frac{\o(\o^2+4\pi^2 T^2)}{6(e^{\o/T}-1)}
.\end{eqnarray}
Replacing the sum over $m_j$ by an integration 
[see Eq.~(\ref{E:KK-summe})]
we can carry out the
$\o$-integration to get
\begin{eqnarray} \label{E:emissivity-approx}
  \frac{d{\cal E}_X}{dt} \label{E:emissivity-approximation}
   &=&
   \frac{p_F^3}{32\pi^7}
   G_N \Omega_d R^d Y_d
   \int_0^\pi d\a\,\sin^3\a 
   \int_0^\pi d\thCM |{\cal A}|^2(\thCM,\bar{p})
   \int_0^1 dx\,x^{d-1} g_X(x,1,\thCM)
,\end{eqnarray}
where $\bar{p}=p_F\sin\a$. The function $Y_d$ is given by
\begin{equation}
  Y_d
  =
  \frac{T^{4+d}}{6}
  \left[4\pi^2 \Gamma(2+d)\zeta(2+d)+\Gamma(4+d)\zeta(4+d)\right]   
,\end{equation}
where $\zeta(x)$ is the Riemann Zeta function.
This reduces the number of integrations to be carried out numerically 
to three.

The temperature dependence of the emissivity in 
Eq.~(\ref{E:emissivity-approximation}) can be easily understood.
The degenerate quarks must have energies that lie within
$\sim T$ of the Fermi surface and are 
contributing one power of $T$ each.
Because of the energy conserving Delta function 
the energy of the graviton is of order $T$ as well.
Each extra dimension contributes one more power of $T$ from
the KK mode.
Because only four out of the five energies are independent, 
the power of $T$ gets reduced by one and we end up with
the $T^{4+d}$ dependence. 

Furthermore, for energy loss due to KK-graviton ($X=h$) 
radiation---the dominant mode---we can replace the
$x$-integration by
\begin{equation}
   \int_0^1 dx\,x^{d-1} g_h (x,1,\thCM)
   = \left\{ 
       \begin{array}{cl}
         (2-2\log{2})\sin^2{\thCM}  & \text{for } d=2 \\
         0.38\sin^2{\thCM} & \text{for } d=3
       \end{array}
     \right.
,\end{equation}
which introduces an error of about 10\% 
but reduces the number of numerical integrations to two.

\section{Results and Comparison to Nucleon-Nucleon Scattering}
\label{S:results}
For quark matter with equal number fractions for the three 
flavors $u$, $d$, and $s$,
$X_u=X_d=X_s=1/3$,
the total emissivity $\dot{{\cal E}}$ can be written as
a sum over possible combinations of two out of
the three flavors in the plasma:
\begin{equation}
  \dot{{\cal E}}
  =
  \sum_{q=u,d,s} \frac{1}{4} \dot{{\cal E}_{qq}} X_q^2 
  + \dot{{\cal E}_{ud}} X_u X_d 
  + \dot{{\cal E}_{us}} X_u X_s 
  + \dot{{\cal E}_{ds}} X_d X_s
,\end{equation}
where the $1/4$ is the symmetry factor
for the
scattering of like quarks.
Then the formula for the emissivity [Eq.~(\ref{E:emissivity-approx})]
can be used to calculate the total emissivity by simply substituting
\begin{equation}
  |{\cal A}|^2
  \longrightarrow
  \frac{1}{3}\left(|{\cal A}_{qq'}|^2+\frac{|{\cal A}_{qq}|^2}{4}\right)
.\end{equation}

Note that for massless quarks the trace over the 
energy-momentum tensor $T_{\mu\nu}$ is zero and so
the KK-dilatons decouple and need not be considered here.
Furthermore, the contribution of the massless 4D-graviton 
can safely be neglected since it is
several orders of magnitude smaller than that for the 
massive KK-gravitons.

\begin{figure}[htb]
  \centerline
    {\epsfxsize=5in \epsfbox{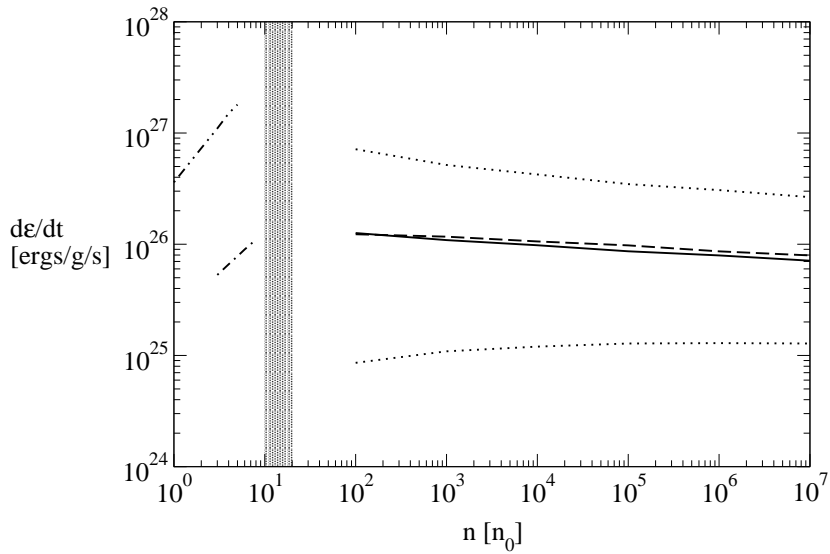}} 
  \caption{Emissivity from quark matter
           for
           high densities of $100-10^7 n_0$.
           The solid [dashed] line shows the result using 
           the formula for the QCD coupling constant derived
           from the 
           $\b$-function to ${\cal O}(\a_s^4)$
           [${\cal O}(\a_s^2)$].
           The cutoff
           mass $m_D$ is calculated using Eq.~(\ref{E:m_D}).
           The upper and lower dotted lines 
           correspond to a cutoff
           mass of $m_D/3$ and $3m_D$, respectively.
           The results for nuclear matter at lower densities
           from 
           Ref.~\protect\cite{HPR00} are also included, using the
           approximations for degenerate (dot-dashed line)
           and non-degenerate matter (dot-dot-dashed line).
           All calculations are for the case of $d=2$ extra dimensions 
           and a temperature of $T=50$~MeV; 
           the size of the extra dimensions is taken to be 1~mm.
           The gray area denotes the region of realistic density
           of $n_q\approx10-20n_0$.}
\label{F:emission2}
\end{figure}

In  Fig.~\ref{F:emission2}
we show the graviton emission rate from quark matter for a density
range of $100-10^7 n_0$ for the case of $d=2$ extra dimensions
and a temperature of $T=50$~MeV,
assuming that the size of the extra dimensions is $R=1$~mm.
The solid line shows the result using the cutoff
mass $m_D$ calculated from Eq.~(\ref{E:m_D}) valid 
for small $\a_s$
and using the formula Eq.~(\ref{E:ashigherorder})
for the QCD coupling constant derived from the $\b$-function
to ${\cal O}(\a_s^4)$.
To estimate the error introduced by
calculating 
only first order diagrams 
we give also the result for the emissivity using the (larger) $\a_s$
[see Eq.~(\ref{E:as})]
calculated from the $\b$-function to ${\cal O}(\a_s^2)$
(dashed line).  
This results in an increase in the cross section which gets partially 
compensated by a larger cutoff mass.  
Therefore the emissivities for both values of $\a_s$ are very similar.

To account for uncertainties in the cutoff mass (higher order corrections,
unaccounted many-body effects)
$m_D$ we also show the emissivity using 
$m_D/3$ (upper dotted line) and
$3m_D$ (lower dotted line).  With $m_D$ spanning about one order
of magnitude in the cutoff mass the emissivity varies by about
1-2~orders of magnitude.
We also present the emission rates from 
nucleon-nucleon scattering~\cite{HPR00}
for lower densities.

Note that it is not sensible to compare the
emission rates for quark matter and nuclear matter
for an equal density, since for densities much higher
than $5 n_0$ nucleons are certainly no longer the 
appropriate degrees of freedom.  Similarly, for densities
smaller than about $100 n_0$ the QCD coupling constant becomes
too large so that perturbative QCD calculations will
no longer be applicable.
It is therefore necessary to extrapolate from the two regions
where reliable calculations are possible to the interesting
transition region with $n_q=10-20n_0$ 
(gray area in Fig.~\ref{F:emission2}).

Determining the emissivity for the physical density region 
from Fig.~\ref{F:emission2} for $d=2$ and carrying out
a similar analysis for $d=3$, 
we can now use the Raffelt criterion, 
limiting the energy lost into
any non-standard physics channel
to $10^{19}$~ergs/g/s, to calculate bounds on the
size of new extra dimensions.
The radius $R$
enters the calculation through the prefactors in 
Eqs.~(\ref{E:emissivity-final}) and (\ref{E:emissivity-approx}).
Our bounds are calculated to be
\begin{eqnarray}
  R&<&2.9\times10^{-4} \text{mm}\quad\text{for }n=2\quad\text{and} \\
  R&<&3.9\times10^{-7} \text{mm}\quad\text{for }n=3
\end{eqnarray}
which are quite similar to the ones determined for nuclear 
matter~\cite{HPR00}.

It should be kept in mind that these bounds become somewhat 
weaker [stronger]
as $m_D$ is taken to larger [smaller] 
values thereby decreasing [increasing] the emissivity.
Our conservative estimates are therefore
weaker than
those found from previous nuclear matter calculations.

\section{Conclusion}
Supernovas like SN1987a provide one of the most stringent constraints
on the size of gravity-only extra dimensions.  
Bounds on these sizes have been 
calculated before~\cite{BHK99,CuP99,HPR00,HPP01} 
assuming a density of 
$\approx 1-3 n_0$ and a typical temperature of $T\approx 50$~MeV
where matter is nucleonic.
The relevant degrees of freedom for this regime are
protons and neutrons. 

In view of the uncertainties with which the parameters 
governing the condition of matter at the star's inner core,
such as density and temperature,
are given, and in view of the fact
that, as can be found from various
nuclear matter equations of state,
matter in the star's core lies very close to a phase transition
to either a deconfined QCD plasma or, for lower 
temperatures, a color superconducting phase, 
it is worthwhile and important
to investigate
how emission rates and bounds on the radius of the 
extra dimensions change
in these regimes.

In this paper we have calculated emissivities and
bounds on the size of gravity-only
extra dimensions from a deconfined quark-gluon plasma in the core
of a star seconds after the supernova core bounce.  We have not
considered the case where the temperature falls below
the critical temperature $T_c$ defining the transition of quark 
matter to a color superconducting state.

By extrapolating the emissivity from a quark-gluon plasma at very high
density ($n_q=100n_0-10^7n_0$) down to the physically interesting region
of $n_q\approx10-20n_0$
we find that the KK-graviton emissivity 
is comparable to those found from 
nuclear matter calculations.  
We have examined the error
introduced into our calculation due to the fact that we are working
only to first order in the QCD coupling constant.
We estimate this error to be at the most 100\%.
To consider the rather
uncertain value of the cutoff mass we have calculated the emissivity
using a cutoff mass which varies about one order of magnitude about
the central value from the lowest order perturbative
calculation [Eq.~(\ref{E:m_D})].
We find that this causes an uncertainty in the emissivity of about
two orders of magnitude.

We have not examined any other many-body effects which could modify
the vacuum rates for KK-graviton emission, such as the  
Landau-Pomeranchuk-Migdal effect, which would decrease
the emissivity because of multiple quark scattering before
the graviton gets emitted.

Note that the source of uncertainties is not the soft radiation 
theorem, which is well suited for the degenerate case, but rather 
our lack of knowledge about the nuclear physics nature of QCD.
This deficit of understanding to only impacts our calculation
of the quark-quark scattering
cross section and the treatment of the QCD plasma,
but also forces us to access the interesting transition region only
with the aid of extrapolation.

Despite the significant uncertainties in our calculation
for quark matter, we believe
to have ruled out the possibility of a significantly larger
emissivity compared to the nuclear matter case. 
It is up to further investigation to provide better information on
the conditions which govern the state of the matter 
at the star's core during and shortly after 
a supernova event.  
It is unlikely that the whole core will consist of dense quark matter,
more likely is a scenario where at the core's center matter is deconfined
and undergoes a phase transition to nuclear matter 
as one goes towards the surface.
Furthermore,
a better understanding of QCD in the non-perturbative regime 
and at finite density would enable a reduction of the uncertainties.

\acknowledgements
I would like to thank Daniel Phillips, Sanjay Reddy, and Martin Savage 
for very helpful discussions and for useful comments on the manuscript.
I also thank
Zacharia Chacko, Jason Cooke, Patrick Fox, Christoph Hanhart, 
and Michael Strickland for interesting
conversations during various stages of this project.  
This work is supported in part by the U.S.\ Department of Energy
under Grant No.\ DE-FG03-97ER4014.
\appendix

\section{Results for the Angular Integration}
The functions $g_X(x,y,\thCM)$ 
defined in Eq.~(\ref{E:g-define}) are calculated 
for the three cases of 4D-graviton ($g$), 
KK-graviton ($h$),
and KK-dilaton ($\phi$) to be
\begin{eqnarray}
  g_g
  &=&
  \frac{1}{4}
  \Biggl\{
    -1+y^2+\frac{(1+6y^2+y^4)\log\frac{y+1}{y-1}}{2y}
    -\frac{2y^2+y^4+4y^2\cos\thCM+\cos 2\thCM}{{\cal W}(x,y,\cos\thCM)} 
                                                    \nonumber \\
    && \quad
     \times
     \log\frac{-(y^2-1)^2+2(y^2+\cos\thCM)^2+2(y^2+\cos\thCM)
                                                {\cal W}(x,y,\cos\thCM)}
              {(y^2-1)^2}
  \Biggr\} \nonumber \\
  &&
  +\frac{1}{4}\Biggl\{\cos{\thCM}\rightarrow -\cos{\thCM}\Biggr\}
,\end{eqnarray}
\begin{eqnarray}
  g_h
  &=&
  \frac{1}{6}\sqrt{1-x^2}
  \Biggl\{
    \frac{4(y^2-1)^2}{x^2+y^2-1}
    +\frac{2(y^4+4y^2+1)\log\frac{y+\sqrt{1-x^2}}{y-\sqrt{1-x^2}}}
          {\sqrt{1-x^2}y} \nonumber \\
    &&\quad
    +\frac{12y^2\cos\thCM +(2y^2+1)^2+3\cos 2\thCM}
          {(x^2-1){\cal W}(x,y,\cos\thCM)} \nonumber \\
    && \qquad 
     \times
     \log
     \frac{(x^2-1)\left[-\frac{(x^2+y^2-1)^2}{(x^2-1)}
                   +2\left(\frac{y^2}{1-x^2}+\cos\thCM\right)^2
                   +2\left(\frac{y^2}{1-x^2}+\cos\thCM\right)
                    {\cal W}(x,y,\cos\thCM)\right]}
          {(x^2+y^2-1)^2}
  \Biggr\} \nonumber \\
  &&
  +\frac{1}{6}\sqrt{1-x^2}\Biggl\{\cos{\thCM}\rightarrow -\cos{\thCM}\Biggr\}
,\end{eqnarray}
and
\begin{eqnarray}
  g_\phi
  &=&
  \frac{d-1}{3(d+2)}\sqrt{1-x^2}(y^2-1)
  \Biggl\{
    \frac{y^2-1}{x^2+y^2-1}
    +\frac{(y^2-1)\log\frac{y+\sqrt{1-x^2}}{y-\sqrt{1-x^2}}}
          {2\sqrt{1-x^2}y}
    +\frac{y^2-1}
          {(x^2-1){\cal W}(x,y,\cos\thCM)} \nonumber \\
    && \quad
     \times
     \log
     \frac{(x^2-1)^2\left[-\frac{(x^2+y^2-1)^2}{(x^2-1)^2}
                   +2\left(\frac{y^2}{1-x^2}+\cos\thCM\right)^2
                   +2\left(\frac{y^2}{1-x^2}+\cos\thCM\right)
                    {\cal W}(x,y,\cos\thCM)\right]}
          {(x^2+y^2-1)^2}
  \Biggr\} \nonumber \\
  &&
  +\frac{d-1}{3(d+2)}\sqrt{1-x^2}(y^2-1)
        \Biggl\{\cos{\thCM}\rightarrow -\cos{\thCM}\Biggr\}
.\end{eqnarray}
The function ${\cal W}(x,y,h)$, 
introduced for convenience, is defined as
\begin{equation}
  {\cal W}(x,y,h)=\sqrt{\frac{(h+1)[-x^2-2y^2+(x^2-1)h+1]}{x^2-1}}
.\end{equation}


\end{document}